\documentclass[draft,jgrga]{AGUTeX}

  \usepackage[dvipdf]{graphicx}

  \setkeys{Gin}{draft=false}
\authorrunninghead{YERMOLAEV ET AL.}

\titlerunninghead{RECOVERY PHASE}

\begin{document}

%
%

\title{Recovery phase of magnetic storms induced by different 
interplanetary drivers
}
%
%

%
%




\authors{Yu. I. Yermolaev, \altaffilmark{1}
I. G. Lodkina, \altaffilmark{1} 
N. S. Nikolaeva , \altaffilmark{1} 
M. Yu. Yermolaev \altaffilmark{1}
}

\altaffiltext{1}{Space Plasma Physics Department, Space Research Institute, 
Russian Academy of Sciences, Profsoyuznaya 84/32, Moscow 117997, Russia. 
(yermol@iki.rssi.ru)}






%
%


\begin{abstract}

Statistical analysis of Dst behaviour during recovery phase of magnetic storms induced 
by different types of interplanetary 
drivers is made on the basis of OMNI data in period 1976-2000. We study storms induced 
by ICMEs (including magnetic 
clouds (MC) and Ejecta) and both types of compressed regions: corotating interaction regions 
(CIR) and Sheaths. The shortest, 
moderate and longest durations of recovery phase are observed in ICME-, CIR-~, and 
Sheath-induced storms, respectively. 
Recovery phases of strong 
($Dst_{min} < -100$ nT) magnetic storms are well approximated by hyperbolic functions 
$Dst(t)= a/(1+t/\tau_h)$ 
with constant $\tau_h$ times for all types of drivers while for 
moderate ($-100 < Dst_{min} < -50$ nT) storms 
$Dst$ profile can not be approximated by hyperbolic function with constant $\tau_h$ because 
hyperbolic time $\tau_h$ increases with increasing time of recovery phase. 
Relation between duration and value $Dst_{min}$ for storms induced by ICME and Sheath has 2 parts:
$Dst_{min}$ and duration correlate at small durations while  they anticorrelate at large durations.

\end{abstract}

%
%

%

\begin{article}

%
%

\section{Introduction}

Dynamics of a magnetic storm is result of competing processes: excitation and relaxation 
\citep{Burtonetal1975}.
Generation of magnetic storm is connected with southward ($Bz < 0$) component of interplanetary magnetic field (IMF)
(or interplanetary electric field  $Ey = Vx \times Bz$, where  $V$ is solar wind velocity) 
\citep{Dungey1961,FairfieldCahill1966,RostokerFalthammar1967,Russelletal1974,Burtonetal1975,Akasofu1981}.
Relaxation of magnetic storm is connected with ring current decay 
\citep{Daglisetal1999}.
The relation between excitation and relaxation changes during storm development: excitation process prevails over relaxation process at the 
main phase while  the relaxation prevails at the recovery phase.

As well known, large and long component $Bz < 0$ in the interplanetary space is observed only in disturbed types of solar 
wind. Therefore interplanetary drivers of magnetic storms are the following disturbed types of solar wind: (1) CIR (Corotating 
Interaction Regions) formed in compress region between slow and high-speed streams of solar wind, (2) ICME (interplanetary 
coronal mass ejections including magnetic clouds and less powerful disturbance - Ejecta), and (3) Sheath formed in compress
 region between fast ICME and slower stream of solar wind (see reviews and recent papers, for instance, by 
\cite{TsurutaniGonzalez1997,Gonzalezetal1999,YermolaevYermolaev2006,Zhangetal2007,Turneretal2009,Yermolaevetal2010c,Yermolaevetal2011,Nikolaevaetal2011,Gonzalezetal2011,Guoetal2011} and references therein).
One of recent important experimental results is evidence that features of magnetic storm depend on type of the 
interplanetary driver 
\citep{BorovskyDenton2006,Dentonetal2006,Huttunenetal2006,Pulkkinenetal2007a,PlotnikovBarkova2007,Longdenetal2008,Turneretal2009,Yermolaevetal2010c,Guoetal2011}
These facts indicate that mechanisms of magnetic storm generation (or modes of these mechanisms) 
can differ depending on the driver. 

It is usually considered that decay of the ring current depends exclusively on internal processes (charge exchange, Coulomb 
collision, wave-particle interaction, and drift loss) and current interplanetary conditions (see, for instance, 
\cite{Daglisetal1999,Keikaetal2006,XuDu2010}. 
and references therein)
In this paper we study time variation of $Dst$ index during recovery phase of magnetic storms induced by different interplanetary 
drivers, i.e. dependence of $Dst$ profile on conditions which were in interplanetary space during the main phase. 

In accordance with formula by 
\cite{Burtonetal1975} 
the $Dst$ index should grow with exponential law at recovery phase. Many papers showed 
that for $Dst$ profile approximation it is necessary to use exponential time ($\tau_e$) which changes with time and depends on 
minimum $Dst$ index \citep{OBrienMcPherron2000,Feldsteinetal2000,MonrealLlop2008,XuDu2010}. 
\cite{Aguadoetal2010} indicated that $Dst$ profile can be approximate by hyperbolic function 
$Dst(t) = Dst_{0}/(1 + t/\tau_h)$ with constant hyperbolic time $\tau_h$. In our analysis we use hyperbolic approximation as simpler 
one because our aim is to search for difference of $Dst$ profile for storms generated 
by various interplanetary drivers and type 
of approximation function is not important for this aim. 

\section{Methods} 

In order to study the problem we use OMNI data of solar wind and interplanetary magnetic field parameters 
(see http://omniweb.gsfc. nasa.gov \citep{KingPapitashvili2004})  and data on $Dst$ index 
(see http://wdc.kugi.kyoto-u.ac.jp/index.html), 
as well as our catalog of large-scale interplanetary events for 
period of 1976--2000 (see ftp://www.iki.rssi.ru/pub/omni  
\citep{Yermolaevetal2009}). 
We study moderate and strong magnetic storms with $-100 < Dst_{min} \le -50$ nT and 
$Dst_{min} < -100$ nT, respectively. 
The technique of determination of connection between 
magnetic storms and their interplanetary drivers consists in the following. 
If the minimum of $Dst$ index lies in an 
interval of a type of solar wind streams or is observed within 1--2 hours after it 
we believe that the given 
storm has been generated by the given type of streams 
\citep{Yermolaevetal2010a}. 
Our analysis for period of 1976--2000 
showed that 145 magnetic storms (31.2\% from total number of identified storms) have been 
generated by CIR, 96 (20.7\%) storms by Sheath, 62 (16.0\%) by MC and 161 (34.7\%) by Ejecta. 
The sources of other 334 magnetic storms (i.e., 42\% of 798 storms observed during this time interval) 
are indeterminate and these 
storms are indicated as IND 
\citep{Yermolaevetal2010a,Yermolaevetal2012}. 
About 20\% of storms were multistep ones during recovery phase and these storms were excluded from analysis. 

It is should be noted that average duration of main phase of magnetic storm is about 7 $\pm$ 4 hours 
and it is less than 
average durations of interplanetary drivers: 24 $\pm$ 11 h for
MC, 29 $\pm$ 5 h for Ejecta, 16 $\pm$ 3 h for Sheath before Ejecta, 9 $\pm$ 5 h for Sheath before MC, 
and 20 $\pm$ 4 h for CIR 
\citep{Yermolaevetal2007c,Yermolaevetal2010a}.
Therefore the main phase for the majority of magnetic storms may be completely defined by one type of 
solar wind streams and it is possible to consider that a storm is connected with one interplanetary 
driver. 
In contrast with main phase, the recovery phase can last up to 3 days, i.e. it is 
longer than duration of interplanetary drivers. 
Therefore  only at initial (short in comparison with full phase) stage of recovery phase 
the interplanetary conditions may correspond to that driver which operated at the main phase, 
and at the subsequent stage of recovery phase type of solar wind streams is replaced by another types: 
Fast follows CIR, MC or Ejecta follow Sheath, Fast or Slow streams follows MC and Ejecta. 
Change of type of solar wind during recovery phase is not analyzed in the given work, and 
we classify storms only on their drivers operating at the main phase of storm.

One of the difficulties of recovery phase analysis is definition of time when the phase comes to its end 
as $Dst$ index quickly grows at the first stage, and then the growth slows down at approach to 
the initial undisturbed level, and there is large (up to 20\%) $Dst$ variation 
which is not connected with general process of storm evolution. 
We use times when $Dst$ index achieves levels of 1/2 and 1/3 from minimum $Dst$ index 
as criteria of time of recovery phase termination, and analyze two durations $\Delta t_{1/2}$ and 
$\Delta t_{1/3}$, i.e. time intervals from $Dst_{mim}$ up to $(1/2)Dst_{mim}$ 
($\Delta t_{1/2} = t((1/2)Dst_{mim}) - t(Dst_{mim})$) and $(1/3)Dst_{mim}$ 
($\Delta t_{1/3} = t((1/3)Dst_{mim}) - t(Dst_{mim})$), respectively. 
Comparison of two data sets corresponding to $\Delta t_{1/2}$ and $\Delta t_{1/3}$ allows us to 
make conclusions about dynamics of $Dst$ index during storm recovery phase.


\section{Results}

Histograms of recovery phases durations $\Delta t_{1/2}$ (black line) and $\Delta t_{1/3}$ (gray line) 
for the storms generated by various drivers are shown in Figure 1. 
All distributions $\Delta t_{1/2}$ have peaks in range of 6--12 hours, some events in range of 0--6 hours
and long tails in range of large times. 
In contrast with $\Delta t_{1/2}$, character of distribution $\Delta t_{1/3}$ 
depends on type of drivers. For Ejecta and MC (and their sum) distributions have a maximum 
in range of 12--18 hours, approximately half of maximum in range of 6--12 hours and 
sharply falling down tail: $\sim 1/10$ of maximum for Ejecta and $\sim 1/3$ for MC at 42 hours. 
Histogram for CIR has a maximum in range of 24--30 hours and wide distribution. 
For Sheath there are 2 maxima at 12--24 hours (this maximum is similar to maximum Ejecta and MC
) and 36--42 hours and wide distribution. 
IND-storms (with not identified sources) have the distribution similar to the sum of mentioned 
above distributions. 
Mean values and standard deviations of durations of recovery phases for both 
$\Delta t_{1/2}$ and $\Delta t_{1/3}$ and for all types of drivers (see Table 1) quantitatively support 
features of Figure 1. 
Thus, both figure and table show that distributions differ for ICME and compressed types of drivers. 
At transition from $\Delta t_{1/2}$ to $\Delta t_{1/3}$ distributions for Ejecta and MC (and their sum) 
are simply displaced towards longer durations with distribution form preservation 
(sharp maximum at small durations and monotonously falling down tail in the range of large durations), 
while CIR- and Sheath-induced storms 
qualitatively change the shape of distributions 
(larger time of maxima and wider disributions).
 
Figure 2 shows $Dst$ profiles during recovery phase of strong ($Dst_{min} <$--100 nT) magnetic 
storms induced by various drivers, and their approximations by hyperbolic functions 
(approximation coefficients are presented in Table 2). 
Data for every type of drivers were obtained by method of superposed epoch analysis 
with zero (epoch) time equal to time of $Dst_{mim}$. 
All profiles are well approximated in region of 0--48 hours, the largest time $\tau_h$ is observed 
for Sheath-induced storms and the smallest $\tau_h$ for ICME-induced storms. 
Unlike strong storms, moderate storms (see Figure 3) for all drivers have a characteristic break 
in range of 10--15 hours, and they cannot be described by hyperbolic function with constant $\tau_h$ 
(or exponential function with constant $\tau_e$). 
For example, we show in Figure 3a, 3b and 3c results of approximation in ranges of 0--11, 0--18 and 
0--48 hours, respectively, and we present approximation coefficients in the Table 2. For all 
approximation regions time $\tau_h$ is the largest for Sheath-induced storms and the least 
for ICME-induced storms.

In order to investigate how duration of storm recovery phase depends on value $Dst_{min}$, 
we calculate average $|Dst_{min}|$ values for each type of drivers in 6-h bins, 
and results are shown in Figure 4. 
Panels 4a and 4b show results for durations $\Delta t_{1/2}$ and $\Delta t_{1/3}$, 
respectively. For all types of storms on the average the size of storms grows in a range 
from 4 till 18 hours for $\Delta t_{1/2}$ durations 
and in a range of 6--30 hours for $\Delta t_{1/3}$, 
and  falls with increasing duration. 
For $\Delta t_{1/3}$
the size of storms is practically a constant for duration longer 48 hour, 
i.e. it does not depend on duration. 
Near maxima $|Dst_{min}|$ stronger storms were induced by Sheath and MC, 
Ejecta-induced storms have moderate peak and CIR-induced storms have weak maximum. 
It should be noted that maxima $|Dst_{min}|$ on both panels 
are statistically significant (see Figure 1). 
Thus, there are 2 branches of $Dst_{min}$ versus duration dependence: increasing and decreasing ones.

\section{Discussion and conclusions} 

On the basis of OMNI data of plasma and IMF parameters of solar wind during 1976-2000 
we classified various types of solar wind streams and found interplanetary drivers for 572 
magnetic storms. These data allowed us to compare temporal evolution of $Dst$ index during 
recovery phase of magnetic storms induced by CIR, Sheath and ICME (including MC and Ejecta). 
Our study allowed to obtain following results. 


1. As distributions of duration of recovery phases show (see Figure 1 and Table 1), 
initial parts of recovery (distributions of $\Delta t_{1/2}$) are close to each other 
for all types of drivers, and at further part of recovery ($\Delta t_{1/3}$) 
distributions differ and histograms are wider 
(i.e. on the average recovery is slower) 
for compression regions Sheath and CIR than for both types of ICME.

2. Dst index during recovery phase is well approximated by hyperbolic functions $Dst(t)= a/(1+t/\tau_h)$ 
only for strong ($Dst_{min} < -100$ nT) storms. This result is in good agreement with conclusions by 
\cite{Aguadoetal2010}.  
Approximation shows that the hyperbolic index $\tau_h$ is the greatest for Sheath-  
(the slowest recovery), intermediate for CIR- and the least for ICME-induced storms (Figure 2 and Table 2). 

3. In contrast to strong storms, Dst index during recovery phase is bad approximated 
by hyperbolic functions for moderate storms with $-100 < Dst_{min} \le -50$ nT and 
hyperbolic index $\tau_h$ depends on time of recovery phase.  
Approximation in regions of 0--11, 0--18, and 0--48 hours (see Figure 3) shows 
that dependence of hyperbolic index 
$\tau_h$ on type of drivers is similar to dependence for strong storms. 

4. For storms generated by ICME and Sheath, there are 2 classes of storms (see Figure 4): 
duration of recovery phase and $Dst_{min}$ correlate for short storm duration 
and they have inverse correlation for long storm duration. 
For storms generated by CIR, duration of recovery phase of storm does not depend on $Dst_{min}$.

Thus obtained results allow us to conclude that recovery phase $Dst$ variations depend on type of 
interplanetary drivers inducing magnetic storms and magnetosphere remember type of driver 
during recovery phase. 


%
%
%
%
%
%
%

\begin{acknowledgments}
The authors are grateful for the possibility of using the OMNI database. The OMNI data were obtained from 
the GSFC/SPDF OMNIWeb on the site http://omniweb.gsfc.nasa.gov. This work was supported by the 
Russian Foundation for Basic Research, projects nos. 07--02--00042 and 10--02--00277a, and by the
 Program 16 of Physics Department of Russian Academy of Sciences (OFN RAN).
\end{acknowledgments}

\end{article}


%
%

%
%
%
%
%


\begin{figure}
\noindent\includegraphics[width=15cm]{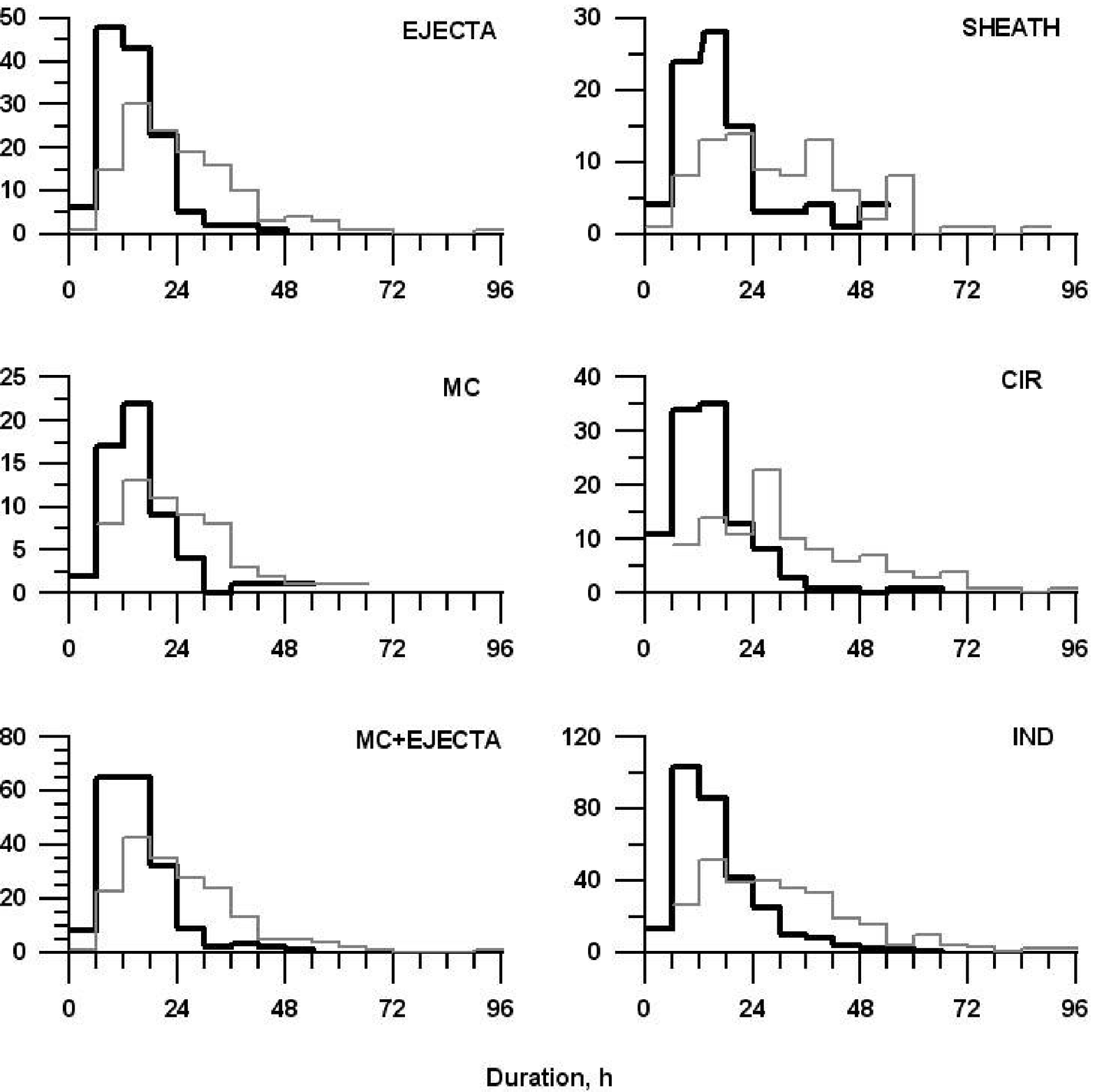}
\caption{Distributions of magnetic storm recovery durations $\Delta t_{1/2}$ (black line) and 
$\Delta t_{1/3}$ (gray line)
for different types of interplanetary drivers 
}
\end{figure}


\begin{figure}
\noindent\includegraphics[width=10cm]{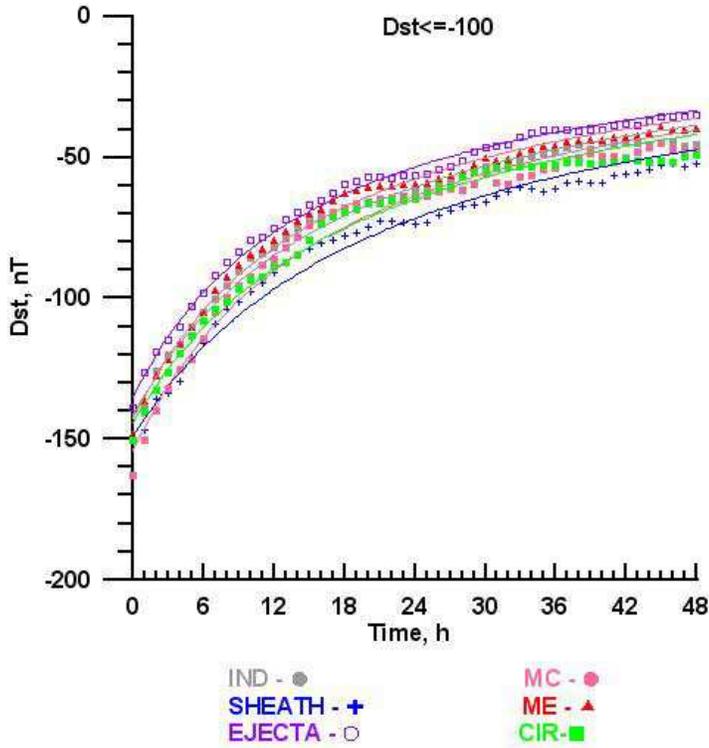}
\caption{Temporal evolution of recovery phases of strong ($Dst_{min} < -100$ nT) storms induced by various types of drivers (points) and hyperbolic approximations  in time intervals of 0--48 hours.}
\end{figure}

\begin{figure}
\noindent\includegraphics[width=10cm]{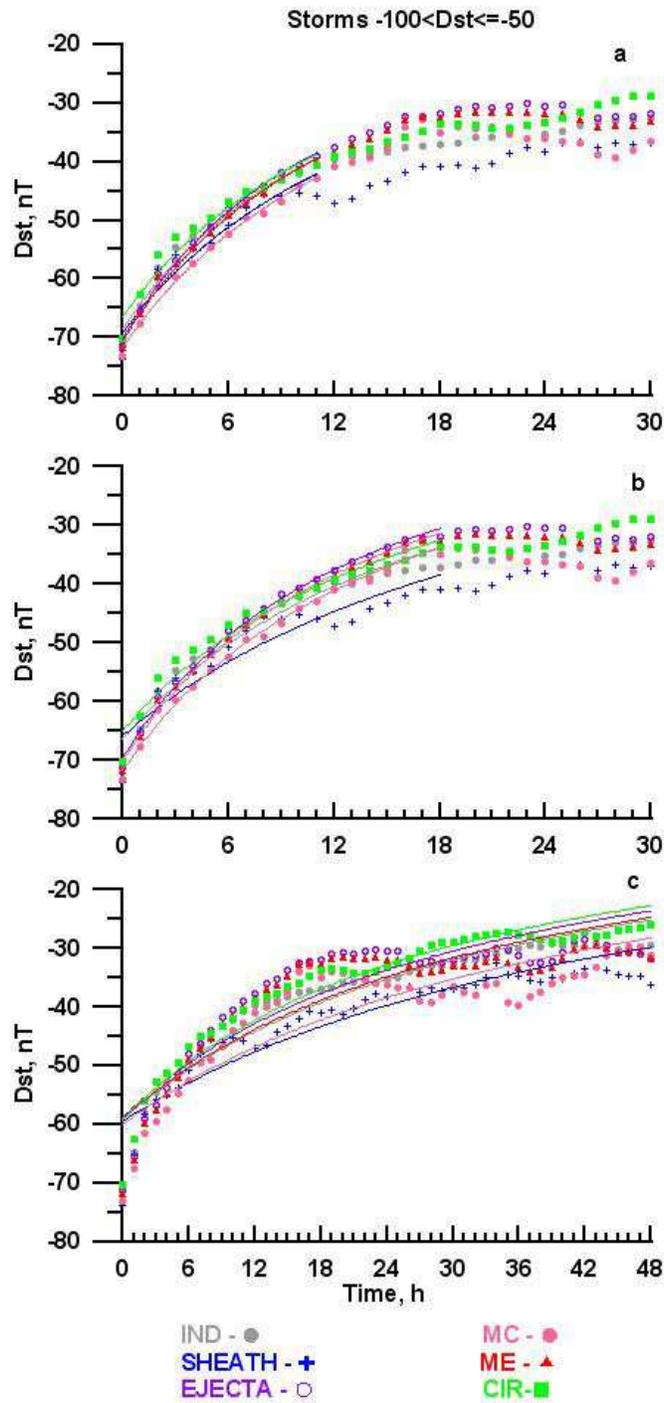}
\caption{Temporal evolution of recovery phases of moderate ($-100 < Dst_{min} < -50$ nT) storms induced by various types of drivers (points) and hyperbolic  approximations  in time intervals of (a) 0--11 hours, (b) 0--18 hours, and (c) 0--48 hours.}
\end{figure}

\begin{figure}
\noindent\includegraphics[width=12cm]{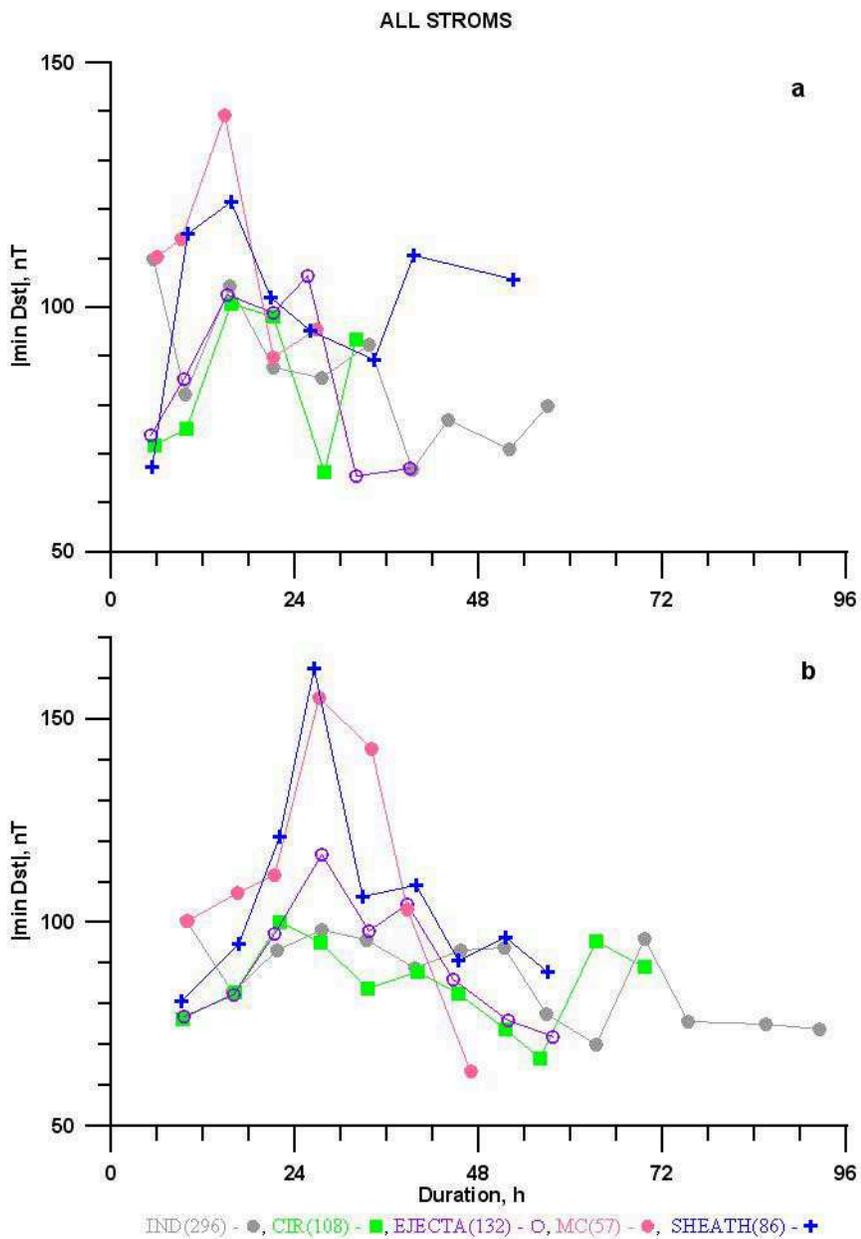}
\caption{Dependence between $Dst_{min}$ and duration of recovery phase of magnetic storms induced by various types of drivers: (a) for $\Delta t_{1/2}$ and (b) for $\Delta t_{1/3}$
}
\end{figure}

%

\begin{table}
\caption{Average values and standard deviations of recovery phase durations.
}
\centering
\begin{tabular}{l| cc|cc}
\hline
             & \multicolumn{2}{c}{$\Delta t_{1/2}$\tablenotemark{a}} & 
               \multicolumn{2}{|c}{$\Delta t_{1/3}$\tablenotemark{b}}  \\
\cline{2-5}
 Types of drivers  & Durations, h & Number & Durations, h & Number \\
\hline
  IND                     & 17.1 $\pm$ 9.9 & 296  & 31.3 $\pm$ 17.0 & 288  \\
  CIR                     & 16.1 $\pm$ 9.8 & 108  & 33.4 $\pm$ 18.0 & 102 \\
  Ejecta                  & 14.8 $\pm$ 7.0 & 130  & 25.9 $\pm$ 14.1 & 128  \\
  MC                      & 16.3 $\pm$ 9.0 &  57  & 25.0 $\pm$ 12.2 &  57  \\
  MC + Ejecta (ME)        & 15.3 $\pm$ 7.7 & 187  & 25.6 $\pm$ 13.5 & 185 \\
  Sheath                  & 18.6 $\pm$ 11.3 & 86  & 31.6 $\pm$ 16.6 &  85  \\
\hline
\end{tabular}
\tablenotetext{a}{$\Delta t_{1/2} = t(1/2 Dst_{min}) - t(Dst_{min})$}
\tablenotetext{b}{$\Delta t_{1/3} = t(1/3 Dst_{min}) - t(Dst_{min})$}
\end{table}

\begin{table}
\caption{Approximation of Dst profiles by hyperbolic function $Dst(t)= a/(1+t/\tau_h)$ \\ for moderate 
         and strong magnetic storms induced by various drivers
}
\centering
\begin{tabular}{l| cc|cc|cc|cc}
\hline
             & \multicolumn{6}{c}{$ -100 \le Dst_{min} <-50 $} & 
               \multicolumn{2}{|c}{$ Dst_{min}<-100$}  \\
\cline{2-9}
	& \multicolumn{2}{c}{$(0-11)h$\tablenotemark{*}} 
        & \multicolumn{2}{|c}{$(0-18)h$\tablenotemark{*}} 
        & \multicolumn{2}{|c}{$(0-48)h$\tablenotemark{*}} 
        & \multicolumn{2}{|c}{$(0-48)h$\tablenotemark{*}} \\
\cline{2-9}
 Types of driver & a, nT & $\tau_h$, h & a, nT & $\tau_h$, h & a, nT & $\tau_h$, h & a, nT & $\tau_h$, h    \\
\hline
IND	&-68.4	&14.97	&-66.3	&18.62	&-59.2	&35.46	&-141.5	&17.95 \\
CIR	&-66.4 	&15.43	&-65.0	&18.05	&-59.2 	&29.94	&-144.1	&19.69 \\
EJE	&-70.1 	&13.39	&-69.6 	&14.05	&-58.8	&32.36	&-135.1	&15.72 \\
MC	&-71.5 	&16.26	&-71.7 	&16.08	&-60.0 	&42.74	&-154.3	&16.95 \\
ME	&-70.4 	&14.05	&-70.1	&14.54	&-59.0	&34.72	&-142.8	&16.23 \\
SHE	&-69.0 	&17.15	&-65.8	&25.25	&-59.5	&48.31	&-148.8	&22.37 \\
\hline
\end{tabular}
\tablenotetext{*}{Time interval of approximation.}
\end{table}


\end{document}